\documentclass[aps,twocolumn,superscriptaddress]{revtex4-1}

\usepackage{graphicx}
\usepackage{color}
\usepackage[normalem]{ulem}
\usepackage{bm}

\begin{document}

\title{Microscopic theory of a superconducting gap in the quasi-one-dimensional organic conductor (TMTSF)$_2$ClO$_4$: Model derivation and two-particle self-consistent analysis}

\author{Hirohito Aizawa}
\email[Electronic address: ]{aizawa@kanagawa-u.ac.jp}
\affiliation{Institute of Physics, Kanagawa University, Yokohama, Kanagawa 221-8686, Japan}

\author{Kazuhiko Kuroki}
\affiliation{Department of Physics, Osaka University, Toyonaka, Osaka 560-8531, Japan}

\date{\today}

\begin{abstract}
We present a first-principles band calculation for the quasi-one-dimensional (Q1D) organic superconductor (TMTSF)$_2$ClO$_4$. 
An effective tight-binding model with the TMTSF molecule to be regarded as the site is derived from a calculation based on maximally localized Wannier orbitals. 
We apply a two-particle self-consistent (TPSC) analysis by using a four-site Hubbard model, which is composed of the tight-binding model and an on-site (intramolecular) repulsive interaction, which serves as a variable parameter. 
We assume that the pairing mechanism is mediated by the spin fluctuation, and the sign of the superconducting gap changes between 
the inner and outer Fermi surfaces, 
which correspond to a $d$-wave gap function in a simplified Q1D model. 
With the parameters we adopt, the critical temperature for superconductivity estimated by the TPSC approach is approximately 1K which is consistent with experiment. 
\end{abstract}

\maketitle

\section{Introduction}

Organic conductors composed of tetramethyltetraselenafulvalene with anion $X$ [(TMTSF)$_{2}X$; $X$ = PF$_{6}^{-}$, AsF$_{6}^{-}$, ClO$_{4}^{-}$, etc.], called "Bechgaard salts"\cite{Bechgaard1980}, have interesting physical properties
\cite{IYT1998,Jerome2004,Dupuis2006,Degiorgi2006-JPSJ-Rev,Lee2006-JPSJ-Rev,Kuroki2006-JPSJ-Rev,Jerome2016}. 
They exhibit a quasi-one-dimensional (Q1D) electronic structure
\cite{Mori1982,Grant1982,Whangbo1982,Ducasse1986,Ishibashi1999} and, in a simple model, their Fermi surface (FS) is a pair of sheets. 
(TMTSF)$_{2}$PF$_{6}$, which has an octahedral anion PF$_{6}$, exhibits spin-density waves (SDWs)
\cite{Bechgaard1980, Scott1980, Mortensen1981} 
or, at ambient pressure, both SDWs and charge-density waves 
\cite{Pouget1996, Kagoshima1999}. 
Under pressure, superconductivity (SC) appears in the vicinity of the SDW phase at 0.9K
\cite{Jerome1980}; the phase diagram for (TMTSF)$_{2}$AsF$_{6}$ also shows a similar relationship between SDW and SC phases
\cite{Brusetti1982}.

Upon slow cooling, (TMTSF)$_{2}$ClO$_{4}$ exhibits SC at 1.2K and ambient pressure
\cite{Bechgaard1981}, 
where an anion-ordering (AO) transition appears at 24K
\cite{Pouget1983} 
because the anion ClO$_{4}^{-}$ is tetrahedral. 
Upon slow cooling, the SDW phase is absent from the pressure--temperature phase diagram; however, (TMTSF)$_{2}$ClO$_{4}$ exhibits an SDW at 6K upon fast cooling
\cite{Takahashi1982} 
or a field-induced SDW in a magnetic field
\cite{McKernan1995}. 
The AO enlarges the unit cell along the $b$ direction, so four TMTSF molecules and two ClO$_{4}^{-}$ anions are contained in the unit cell 
\cite{Pouget1983}. 
In reciprocal space, the energy-band structure is folded along the $b^{*}$ direction, and the FS consists of two pairs of sheets
\cite{LePevelen2001, Nagai2011, Alemany2014}.

The shape of the theoretical 
FS of (TMTSF)$_{2}$ClO$_{4}$ in the AO state depends on the band-calculation method used to derive it. 
In an early result from an extended H{\"u}ckel-band calculation, the outer FS splits from the inner FS because the site-energy difference between two independent TMTSF molecules is taken to be about 100 meV
\cite{LePevelen2001}. 
Conversely, a first-principles band calculation based on density-functional theory (DFT) shows that the outer and inner FSs are almost in contact 
\cite{Nagai2011, Alemany2014}. 
This implies that the site-energy difference due to the AO is small 
\cite{Nagai2011}; 
it is estimated to be 14 meV
\cite{Alemany2014}. 
Several experiments have estimated the site-energy difference between the two independent TMTSF molecules: 
Assuming the Fermi energy is taken as 0.1 meV, 
the rapid oscillations of the magnetoresistance indicate site-energy difference of 4.5 meV
\cite{Uji1996}. 
These estimates are based on the third angular effect of magnetoresistance and report a site-energy difference of about 0.083$t_{a}$
\cite{Yoshino1999}
later corrected to  0.028$t_{a}$
\cite{Yoshino2003}, 
where $t_{a}$ is the transfer integral along the most conducting $a$ axis. 
Theoretical results that are consistent with the experimental data indicate that the site-energy difference is 0.2$t_{b}$ under assumptions of $t_{a}/t_{b}=10$ 
\cite{Lebed2005} 
or $t_{a}/t_{b}=9.75$ 
\cite{Ha2006}. 
These estimated values are of the order of 10 meV.

For (TMTSF)$_{2}$ClO$_{4}$ at low temperature, the upper critical field $H_{\rm c2}$ obtained from the onset of the critical temperature exceeds the Pauli paramagnetic limit,  therefore the spin-singlet, spin-triplet and Fulde--Ferrell--Larkin--Ovchinnikov states have been discussed\cite{Oh2004, Kusaba2008, Yonezawa2008, Yonezawa2008a}. 
However, several experiments find $H_{\rm c2}$ to be at the Pauli paramagnetic limit
\cite{Murata1987, Yonezawa2012}. 
Although the SDW phase is absent in the pressure--temperature phase diagram of relaxed (TMTSF)$_{2}$ClO$_{4}$, the possibility of an anisotropic SC gap has been suggested by several experiments, such as experiments that investigated the impurity effect
\cite{Coulon1982,Tomic1983,Joo2004,Joo2005} 
and measurements of the NMR relaxation rate $T_{1}^{-1}$. 
For (TMTSF)$_{2}$ClO$_{4}$, these latter measurements show no coherence peak
\cite{Takigawa1987,Shinagawa2007}. 
A recent angle-resolved heat capacity measurement suggests that the SC gap has nodes on the FS\cite{Yonezawa2012}. 
Theoretical studies suggest that, to explain the anisotropic SC gap function, the change of sign at the FS is a good candidate for (TMTSF)$_{2}X$
\cite{Hasegawa1987,Takigawa2006}. 
Conversely, the measurement of the thermal conductivity in (TMTSF)$_{2}$ClO$_{4}$ reveals a nodeless SC gap but does not exclude unconventional SC
\cite{Belin1997}. 
Focusing on the splitting of the FS in (TMTSF)$_{2}$ClO$_{4}$, Shimahara suggested a nodeless $d$-wave gap in which the line nodes lie between the outer and inner FSs (interFSs)
\cite{Shimahara2000}. 
Recently, muon-spin rotation ($\mu$SR) measurements suggested odd-frequency spin-singlet $p$-wave pairing as the bulk SC state
\cite{Pratt2013}. 
In addition, a theoretical work that assumes a pairing mechanism mediated by spin fluctuations that coexist with charge fluctuations suggested odd-frequency spin-singlet $p$-wave pairing in the extended Hubbard model applied to a Q1D system
\cite{Shigeta2013}.

Previous studies based on a model for (TMTSF)$_{2}$ClO$_{4}$ that involves four FSs suggested several pairing states 
\cite{Shimahara2000,Mizuno2011,Nagai2011,Alemany2014,Pratt2013,Yonezawa2012,Miyawaki2016}, 
such as a nodeless $d$-wave, nodal $d$-wave, and nodal $g$-wave states. 
Note that the "symmetry" of the $d$($p$)-wave and $s$($f$)-wave gaps are the same in the system on which we focus herein; however, we call these states $d$($p$)-wave and $s$($f$)-wave gaps in the broad sense, meaning that the sign of the gap changes along the FS. 
Recently, we showed that the gap function for spin-singlet $d$-wave pairing changes sign between the inerFSs for (TMTSF)$_{2}$ClO$_{4}$\cite{Aizawa2016}. 
Several studies have used the Q1D model for (TMTSF)$_{2}X$ to discuss the anisotropy of the pairing gap, the relationship between SC and SDWs, and physical properties near the quantum critical point
\cite{Hasegawa1986a,Hasegawa1987,Duprat2001,Shahbazi2015}. 
As for the pairing state, the spin-fluctuation-mediated mechanism suggests spin-singlet pairing with a $d$-wave-like gap function 
\cite{Shimahara1989,Kino1999,Kuroki1999,Nomura2001,Fuseya2005a}. 
And another glue, such as electron--phonon interactions and charge fluctuations, which coexists with spin-fluctuation suggests spin-triplet pairing 
\cite{Kohmoto2001,Kuroki2001,Fuseya2002,Kuroki2004,Onari2004,Suginishi2004,Tanaka2004,Fuseya2005,Kuroki2005,Nickel2005}. 
In strongly 1D system, the odd-frequency pairing state has been suggested
\cite{Shigeta2009,Shigeta2011,Shigeta2013}. 
The magnetic-field effect on the pairing competition has been studied using phenomenological or microscopic approaches
\cite{Miyazaki1999,Lebed2000,Shimahara2000a,Belmechri2007,Aizawa2008,Belmechri2008,Aizawa2009,Aizawa2009b,Kajiwara2009,Croitoru2014,Shahbazi2017}. 
Finally, pairing mechanisms that are not mediated by spin or charge fluctuation suggest that anisotropic pairing states may arise \cite{Podolsky2004a,Podolsky2004,Ohta2005}.

In the first part of this paper, we show (i) the electronic band structure and FS obtained from a first-principles band calculation and (ii) the tight-binding model and transfer energies derived from maximally localized Wannier orbitals (MLWOs). 
Next, assuming a pairing mechanism mediated by spin fluctuations, (iii) we discuss the pairing-gap symmetry and its driving force for spin singlet and spin triplet with even- and odd-frequency channels and (iv) estimate the critical temperature $T_{\rm c}$ in the Hubbard model based on the MLWO tight-binding model for (TMTSF)$_{2}$ClO$_{4}$ by applying the two-particle self-consistent (TPSC) method
\cite{Vilk1997}. 

This paper is organized as follows: 
In Sec. \ref{sec-method}, we describe the parameter sets used in the DFT and MLWO calculations, after which we introduce the TPSC method and SC for the multisite Hubbard model. 
In Sec. \ref{sec-result-band-sc}, we present the band structure obtained from first-principles band calculations and the transfer energies of the tight-binding model derived from the MLWO calculation. 
Next, the pairing-gap functions and their possible role in (TMTSF)$_{2}$ClO$_{4}$ are discussed. 
Finally, Sec. \ref{ssec-conclusion} contains the conclusion.

\section{Method}\label{sec-method}

\subsection{Band structure and the effective model}\label{ssec-band-model}

We present a first-principles band calculation based on the all-electron full potential linearized augmented plane-wave (LAPW) + local orbitals method within {\sc wien2k} 
\cite{Blaha2014}. 
This code implements DFT 
\cite{Hohenberg1964,Kohn1965} 
with different possible approximation for the exchange correlation potential, which is calculated using the generalized gradient approximation (GGA) 
\cite{Perdew1996}. 
To attain convergence in the eigenvalue calculation, the single-particle wave functions in the interstitial region are expanded by plane waves with a cutoff of $R_{\rm MT} K_{\rm max}=3.0$, where $R_{\rm MT}$ denotes the smallest muffin-tin radius and $K_{\rm max}$ is the maximum $K$ vector in the plane-wave expansion. 
For (TMTSF)$_{2}$ClO$_{4}$ with the AO, the muffin-tin radii are taken to be 1.31, 1.31, 1.75, 1.24, and 0.67 in atomic units (au) for  Cl, O, Se, C, and H, respectively. 
Thus, $K_{\rm max} = 3.0/0.67 = 4.5$, and the plane-wave cutoff energy is 272.8 eV. 
The self-consistent calculations use 10$\times$4$\times$5 $k$ points in the irreducible Brillouin zone. 

For comparison, the band calculation was done using the {\sc Quantum ESPRESSO} 
code (QE)\cite{Giannozzi2009}, 
which is based on DFT 
\cite{Hohenberg1964,Kohn1965} 
using a plane-wave basis set and pseudopotentials. 
We adopt the GGA 
\cite{Perdew1996} 
and norm-conserving pseudopotentials 
\cite{Troullier1991}. 
The plane-wave cutoff energy is 60 Ry. 
The self-consistent calculations were done using 10$\times$10$\times$10 $k$ points in the irreducible Brillouin zone. 
Next, we derive the tight-binding model with four sites per unit cell by applying MLWOs 
\cite{Marzari1997,Souza2001,Mostofi2008} 
on each TMTSF molecule. 
In the DFT calculation, 
we use the crystal structure measured experimentally at 7K 
\cite{Gallois1987} 
and do not relax the atomic positions. 
We ignore the spin-orbit interaction in the DFT calculations.

\subsection{Many-body effect and superconductivity}\label{ssec-tpsc-sc}

We introduce the Hubbard Hamiltonian $H$ based on the four-site tight-binding model: 
\begin{eqnarray}
 H=\sum_{\left< i \alpha: j \beta \right>, \sigma}
  \left\{ t_{i \alpha: j \beta} 
   c_{i \alpha \sigma}^{\dagger} c_{j \beta \sigma} + {\rm H. c.} 
  \right\} 
  +\sum_{i \alpha} U
  n_{i \alpha \uparrow} n_{i \alpha \downarrow}, 
  \label{Hij}
\end{eqnarray} 
where $i$ and $j$ are unit-cell indices, $\alpha$ and $\beta$ specify the sites in a unit cell, $c_{i \alpha \sigma}^{\dagger}$ ($c_{i \alpha \sigma}$) is the creation (annihilation) operator for spin $\sigma$ at site $\alpha$ in unit cell $i$, $t_{i \alpha: j \beta}$ is the electronic transfer energy between site $(i, \alpha)$ and site $(j, \beta)$, and $\left< i \alpha: j \beta \right>$ represents the summation over bonds that corresponds to the transfer. 
$U$ is the on-site interaction, and $n_{i \alpha \sigma}$ is the number operator for electrons with spin $\sigma$ on site $\alpha$ in unit cell $i$. 
Because we focus on a material configured as $D_2 X$ (where $D$ is the donor molecule, and $X^{-1}$ is the anion), the band is $1/4$ filled in the hole representation (i.e., $3/4$ filled in the electron representation).

To deal with the effect of the electron correlation, we prepare the bare susceptibility and bare Green's function for the site representation given by 
\begin{eqnarray}
 \chi^{0}_{\alpha \beta} \left( q \right)
 &=&-\frac{T}{N_c} \sum_{ k }
  G^{0}_{\alpha \beta }\left( k+q \right) G^{0}_{\beta  \alpha}\left( k \right),   
 \label{chi0}
\\
 G^{0}_{\alpha \beta} \left( k \right)
 &=& \sum_{\gamma}
 d_{\alpha \gamma}\left( \textbf{\textit{k}} \right) 
 d_{\beta \gamma}^{*}\left( \textbf{\textit{k}} \right) 
G^{0}_{\gamma} \left( k \right), 
 \label{g0}
\end{eqnarray}
where $T$ and $N_c$ are the temperature and the total number of unit cells, respectively, $G^{0}_{\gamma} \left( k \right)$ is the bare Green's function in the band representation, and $d_{\alpha \gamma}\left( \textbf{\textit{k}} \right)$ is the unitary matrix. 
Here we introduce the abbreviations 
$k=\left( \textbf{\textit{k}}, i \varepsilon_{n} \right)$ 
and 
$q=\left( \textbf{\textit{q}}, i \omega_{m} \right)$ 
for the fermionic and bosonic Matsubara frequencies, respectively. 
The indices $\alpha\beta$ refer to element ($\alpha$ $\beta$) of matrices such as $\hat{\chi}^{0}\left( q \right)$.

TPSC has been applied to single-site systems
\cite{Vilk1997,Otsuki2012}, 
multisite systems, 
\cite{Arya2015, Ogura2015, Aizawa2015, Aizawa2016}
and multi-orbital systems 
\cite{Miyahara2013}. 
Within the TPSC, and by using the bare susceptibility given by Eq. (\ref{chi0}), the spin and charge susceptibilities are 
\begin{eqnarray}
 \hat{\chi}^{\rm sp}\left( q \right)
  &=&\left[ \hat{I}-\hat{\chi}^{0}\left( q \right) \hat{U}^{\rm sp} \right]^{-1}
   \hat{\chi}^{0}\left( q \right),
  \label{chisp-tpsc} 
  \\
 \hat{\chi}^{\rm ch}\left( q \right)
  &=&\left[ \hat{I}+\hat{\chi}^{0}\left( q \right) \hat{U}^{\rm ch} \right]^{-1}
   \hat{\chi}^{0}\left( q \right), 
  \label{chich-tpsc}
\end{eqnarray}
where $\hat{U}^{\rm sp}$ ($\hat{U}^{\rm ch}$) is the local spin (charge) vertex, and $\hat{I}$ is the unit matrix. 
The local vertices are determined by the two sum rules for the local moment: 
\begin{eqnarray}
 \frac{2T}{N_c}\sum_{q} \chi^{\rm sp}_{\alpha \alpha} \left( q \right) 
 &=& n_{\alpha}-2\left< n_{\alpha \uparrow} n_{\alpha \downarrow} \right>, 
 \label{chisp}
\end{eqnarray}
and
\begin{eqnarray}
 \frac{2T}{N_c}\sum_{q} \chi^{\rm ch}_{\alpha \alpha} \left( q \right)
 &=& n_{\alpha}+2\left< n_{\alpha \uparrow} n_{\alpha \downarrow} \right>
 -n_{\alpha}^{2}, 
 \label{chisp}
\end{eqnarray}
where $n_{\alpha}$ is the particle number at site $\alpha$. 
Here, we have used the relations $n_{\alpha \uparrow} = n_{\alpha \downarrow} = n/2$ and $n_{\alpha \sigma} = n_{\alpha \sigma}^{2}$ from the Pauli principle.

The local spin vertex $\hat{U}^{\rm sp}$ is related to the double occupancy $\left< n_{\alpha \uparrow} n_{\alpha \downarrow} \right>$ by the ansatz 
\begin{eqnarray}
 U^{\rm sp}_{\alpha \alpha} 
 =\frac{ \left< n_{\alpha \uparrow} n_{\alpha \downarrow} \right> }
 { \left< n_{\alpha \uparrow} \right> \left< n_{\alpha \downarrow} \right> }
 U_{\alpha \alpha}, 
 \label{Usp-tpsc}
\end{eqnarray}
where $U_{\alpha \alpha}$ is element ($\alpha$ $\alpha$) of the on-site interaction matrix $\hat{U}$. 
Equation (\ref{Usp-tpsc}) breaks the particle-hole symmetry and should be used for $n_{\alpha} \le 1$.  
When $n_{\alpha} > 1$, the particle-hole transformation is used, and the double occupancy 
$D_{\alpha} = \left< n_{\alpha \uparrow} n_{\alpha \downarrow} \right> $ is given by 
\begin{eqnarray}
D_{\alpha}
=\frac{U^{\rm sp}_{\alpha \alpha} }{ U_{\alpha \alpha} }
\frac{n_{\alpha}^2 }{ 4 }
+\left( 1-\frac{U^{\rm sp}_{\alpha \alpha} }{ U_{\alpha \alpha} } \right)
\left( n_{\alpha}-1 \right)
\theta \left( n_{\alpha}-1 \right), 
 \label{doubocc-tpsc}
\end{eqnarray} 
where $\theta \left( x \right)$ is the step function. 
Equations (\ref{chisp-tpsc})--(\ref{doubocc-tpsc}) give a set of the self-consistent equations for the TPSC method. 
Given $\hat{U}_{\rm sp}$ and $\hat{U}_{\rm ch}$, the interaction for the self-energy takes the form 
\begin{eqnarray}
 \hat{V}^{\Sigma}\left( q \right) = 
 \frac{1}{2} 
 \left[ \hat{U}^{\rm sp} \hat{\chi}^{\rm sp}\left( q \right) \hat{U}  
+\hat{U}^{\rm ch} \hat{\chi}^{\rm ch}\left( q \right) \hat{U}
 \right]. 
 \label{int_selfene}
\end{eqnarray}
From Eq. (\ref{int_selfene}), the self-energy is 
\begin{eqnarray}
 \Sigma_{\alpha \beta} \left( k \right)
 = \frac{T}{N_c}
 \sum_{q} V^{\Sigma}_{\alpha \beta}\left( q \right) 
 G_{\alpha \beta}\left( k-q \right), 
 \label{selfene}
\end{eqnarray}
and the dressed Green's function is 
\begin{eqnarray}
 \hat{G}\left( k \right) 
  &=&\left[ \hat{I}-\hat{G}^{0}\left( k \right)
                    \hat{\Sigma} \left( k \right) \right]^{-1}
   \hat{G}^{0}\left( k \right). 
 \label{gr}
\end{eqnarray}

Assuming a pairing mechanism mediated by spin fluctuation, the pairing interactions for the spin-singlet (SS) and spin-triplet (ST) channels are 
\begin{eqnarray}
 \hat{V}^{\rm SS}\left( q \right) 
  &=& \hat{U}
+\frac{3}{2}\hat{U}^{\rm sp} \hat{\chi}^{\rm sp}\left( q \right) \hat{U} 
-\frac{1}{2}\hat{U}^{\rm ch} \hat{\chi}^{\rm ch}\left( q \right) \hat{U}, 
  \label{pairing_int_ss}
\\
 \hat{V}^{\rm ST}\left( q \right) 
  &=& 
-\frac{1}{2}\hat{U}^{\rm sp} \hat{\chi}^{\rm sp}\left( q \right) \hat{U}
-\frac{1}{2}\hat{U}^{\rm ch} \hat{\chi}^{\rm ch}\left( q \right) \hat{U},
  \label{pairing_int_st}
\end{eqnarray}
respectively. 
By using the obtained pairing interaction, we solve the linearized Eliashberg equation to obtain the transition temperature $T_{\rm c}$ and the SC gap function. 
The linearized Eliashberg equation is given by 
\begin{eqnarray}
 \lambda^{\mu} \varphi^{\mu}_{\alpha \beta} \left( k \right) 
  &=&
  \frac{-T}{N_c} 
  \sum_{ k' \alpha' \beta'}
  V^{\mu}_{\alpha \beta} \left( k-k' \right)
  \nonumber \\
  &\times&
  G_{\alpha \beta'}\left( k' \right)
  G_{\beta \alpha'}\left(-k' \right)
  \varphi^{\mu}_{\alpha' \beta'}\left( k' \right), 
  \label{gap-eq}
\end{eqnarray}
where $\mu$ represents the pairing state, $\lambda^{\mu}$ is the eigenvalue, and $\varphi^{\mu}_{\alpha \beta}\left( k \right)$ is element ($\alpha$ $\beta$) of the gap-function matrix. 
The critical temperature $T_{\rm c}$ is the temperature where $\lambda^{\mu}$ reaches unity. 
For a temperature regime higher than $T_{\rm c}$, we use $\lambda^{\mu}$ as a measure of $T_{\rm c}$.

In the present study, we show the spin susceptibility obtained from the larger eigenvalue of the matrix. 
In the present calculation, we take the system size to be $64 \times 32$ $k$ meshes and $16384$ Matsubara frequencies.

\section{Band structure and superconductivity}\label{sec-result-band-sc}

\subsection{Band structure and the tight-binding model}\label{ssec-res-band-model}

\begin{figure}[!htb]
\begin{center}
\includegraphics[width=5.4cm]{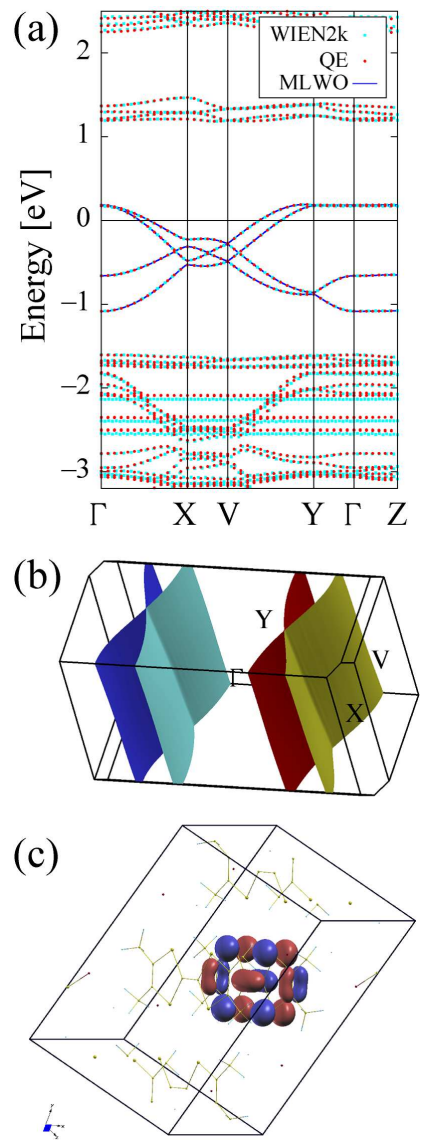}
\end{center}
\caption{
(a) Band structures obtained from the first-principles calculation, where the light blue (red) dotted curve represents the result of {\sc wien2k} (QE) and the Fermi level is taken as zero energy. 
The band structure of the tight-binding model shown in Fig. \ref{fig2} derived from the MLWOs represents the blue solid curves. 
(b) FS for the AO of (TMTSF)$_{2}$ClO$_{4}$ at 7K and 
(c) typical Wannier orbital on the TMTSF molecule, where the red (blue) surfaces indicate positive (negative) isosurface, drawn by {\sc XcrysDen}\cite{Kokalj1999}. 
}
\label{fig1} 
\end{figure}

The first-principles band structures obtained with {\sc wien2k} and QE are almost the same, as shown in Fig. \ref{fig1}(a). 
For both results, it can be seen that the highest-occupied molecular orbital (HOMO) is isolated from the lowest-unoccupied molecular orbital (LUMO). 
Similarly, the HOMO$-$3 is isolated from the lower-energy bands. 
The four band structures near the Fermi level, which is taken as zero energy, are isolated from the other bands. 
Considering the number of donor molecules in a unit cell, we treat the four bands as targets to derive an effective tight-binding model. 
From the band structure along the $k$-path from $\Gamma=(0, 0, 0)$ to ${\rm Z}=(0, 0, \pi / c)$, we find that the three dimensional nature of the electronic structure is weak. 
Hence, the FS, which is composed of four sheets on the $k_z$=0 and $k_z=\pi/c$ planes, is almost same as shown in Fig \ref{fig1}(b). 
Although the dispersion along the $k_z$ direction is important to understand some SC properties, we focus on the two dimensional conducting plane to reveal the SC gap symmetry in this study. 
Two FSs almost touch, which is the same as what was found in previous studies 
\cite{Nagai2011,Alemany2014}, 
and the difference of the site energy between TMTSF chains A and B is non-zero, which is again the same as the previous result
\cite{Alemany2014} 
(we show details later). 

\begin{figure}[!htb]
\begin{center}
\includegraphics[width=7.4cm]{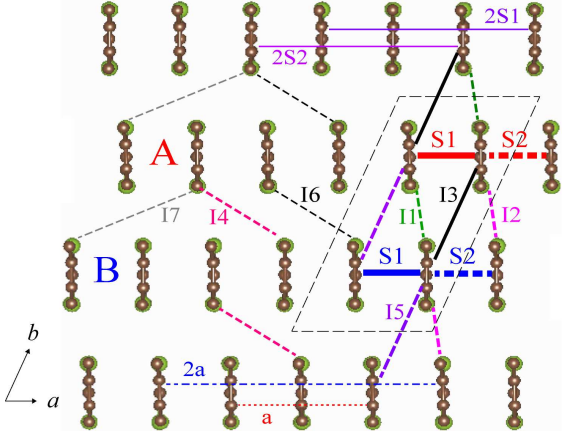}
\end{center}
\caption{
The tight-binding model, where a TMTSF molecule is regarded as a site, and definition of the transfer energies listed in Table \ref{transfers} in the conductive $a$-$b$ plane. 
Note that $t_{S1}$ is the transfer within the dimerized molecules. 
The dashed parallelogram represents the unit cell, and the color of lines corresponds to that of the transfer-energy characters. 
The character of red colored "A" (blue colored "B") represents the TMTSF chain A (B) along the $a$ direction. 
}
\label{fig2} 
\end{figure}

Wannier orbitals are localized on each TMTSF molecule, as shown in Fig. \ref{fig1}(c). 
The band structure derived from the MLWO accurately reproduces the first-principles band structures near the Fermi level, as shown in Fig. \ref{fig1}(a). 
In Table \ref{transfers}, we summarize the difference in site energies and the nearest- and next-nearest-neighboring transfer energies in the tight-binding model where a TMTSF molecule is regarded as a site [see Fig. \ref{fig2}]. 
From the MLWO calculation, we conclude that the site-energy difference $E_{\rm AO}$ is 8.7 meV, whose order of the energy is consistent with the previous studies 
\cite{Uji1996,Yoshino2003,Lebed2005,Ha2006,Alemany2014}. 
From now on, we regard this value as the AO potential. 
The transfer energies along the TMTSF chains, $t_{S1}$ and $t_{S2}$ for  the chains A and B, are much close to the values obtained from the DFT calculation of (TMTSF)$_{2}$ClO$_{4}$ without the AO 
\cite{Ishibashi1999} 
and with the AO 
\cite{Nagai2011,Alemany2014}. 
The previous DFT calculation without the AO\cite{Ishibashi1999} indicates two band structures near the Fermi level because of the dimerization. 
Conversely, in the results with the AO, the dimerization as well as the band folding in the $b^*$ direction bring the four band structures. 
We expect that the AO may affect the transfer energies within the TMTSF chains. 
Namely, the transfers $t_{S1}$ and $t_{S2}$ in the TMTSF chain A differ between them in the chain B, as shown in Fig. \ref{fig2} and Table \ref{transfers}, in addition to the site-energy difference. 
The transfer energies presented by the previous DFT studies also indicate similar nature of the transfer energies
\cite{Nagai2011}. 
From the lengths between the TMTSF and ClO$_{4}$ molecules, effects of the AO on the transfers has been discussed 
\cite{Alemany2014}.

\begin{table}[!htb]
\centering
\caption{Site energies and transfer energies 
in meV for (TMTSF)$_2$ClO$_4$ at 7 K. } 
\begin{tabular}{ l|c c } \hline
\hspace{1pt} $E$ [meV] \hspace{1pt} & 
\hspace{0pt} TMTSF A   \hspace{0pt} &
\hspace{2pt} TMTSF B   \hspace{2pt}               \\ \hline \hline
  $E_{\rm AO}$ &               0  &  8.7          \\ 
  $t_{S1}$          &             271  &  257     \\
  $t_{S2}$          &             255  &  253     \\
  $t_{I1}$          &  \hspace{2cm}   -34         \\
  $t_{I2}$          &  \hspace{2cm}   -72         \\
  $t_{I3}$          &  \hspace{2cm}    55         \\
  $t_{I4}$          &  \hspace{2cm}    -4         \\
  $t_{I5}$          &  \hspace{2cm}    57         \\
  $t_{I6}$          &  \hspace{2cm}    -3         \\ 
  $t_{a }$          &              10  &    9     \\
  $t_{2S1}$         &               3  &    3     \\
  $t_{2S2}$         &               3  &    2     \\
  $t_{2a}$          &               2  &    2     \\
  $t_{I7}$          &  \hspace{2cm}    -2         \\ \hline 
 \end{tabular}
 \label{transfers}
\end{table}

\subsection{Spin susceptibility and the superconducting-gap function}\label{ssec-res-spin-sc}

\begin{figure}[!htb]
\begin{center}
\includegraphics[width=8.4cm]{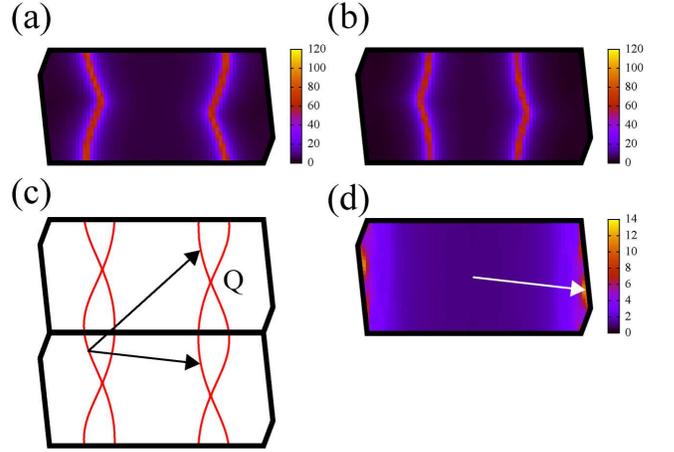}
\end{center}
\caption{
Green's function for (a) outer and (b) inner bands. 
(c) FS and (d) diagonalized spin susceptibility, where arrows represent the nesting vector $\bm{Q}_{2k_{\rm F}}$. 
} 
\label{fig3} 
\end{figure}

We now show the results of the TPSC scheme applied to the multisite Hubbard model for (TMTSF)$_{2}$ClO$_{4}$. 
The bandwidth $W$ is about 1.27 eV, so we take the on-site interaction $U=1.3$ eV to be nearly the same as the bandwidth. 
The on-site interaction $U$ is estimated from other strongly correlated organic conductors by applying the extended H{\"u}ckel calculation 
\cite{Castet1996} 
and  first-principles calculations 
\cite{Cano-Cortes2007,Nakamura2009,Nakamura2012}. 
Based on recent results 
\cite{Cano-Cortes2007,Nakamura2009,Nakamura2012}, 
we consider that the on-site interaction that we have used is appropriate. 
At a temperature of $T=0.002$ eV, 
Fig. \ref{fig3}(a) [\ref{fig3}(b)] show the absolute value of the Green's function for the outer (inner) band, which takes a large value near the FS 
shown in Fig. \ref{fig3}(c). 
From now on, we call the touching points of the outer and inner FSs Q points following the previous studies 
\cite{Pratt2013,Alemany2014}. 
The $k_{x}$ component of the Fermi wavenumber $k_{\mathrm F}$ takes the value $k_{\mathrm F} \approx \pi/2$. 
The diagonalized spin susceptibility is shown in Fig. \ref{fig3}(d). 
The wavenumber at which the spin susceptibility is maximized corresponds to the nesting vector, which is represented by the arrows in Fig. \ref{fig3}(d). 
When the $k_x$ component of the nesting vectors takes the wavenumber $2k_{\rm F}$, we call this the nesting vector $\bm{Q}_{2k_{\rm F}}$. 
As seen in Fig. \ref{fig3}(c), we can verify that the spin susceptibility reaches its maximum at $\bm{Q}_{2k_{\rm F}}$ between the interFSs.

The SC states we considered in this study are both the singlet and triplet channels in spin space, in addition to the even- and odd-frequency pairings
\cite{Berezinskii1974,Balatsky1992,Abrahams1993,Coleman1994,Vojta1999,Fuseya2003,Tanaka2007b,Yanagi2012,Shigeta2012}. 
Adopting the notation of the pairing states from previous studies
\cite{Tanaka2007a,Tanaka2007,Shigeta2011}, 
we label the pairing states by frequency, spin, parity of the gap function, and gap symmetry in wavenumber space, such as even-frequency, spin-singlet, even-parity (ESE), then the gap symmetry is added from the calculation result. 
Similarly, we abbreviate odd-frequency, spin-singlet, odd-parity as OSO; even-frequency, spin-triplet, odd-parity as ETO; and odd-frequency, spin-triplet, even-parity as OTE.

To satisfy the Pauli principle for exchanging electrons, the gap function for spin-singlet pairing must satisfy  
\begin{eqnarray}
   \varphi^{\rm SS} \left( \bm{k}, i\varepsilon_{n} \right) 
 = \varphi^{\rm SS} \left(-\bm{k}, i\varepsilon_{n} \right)
 =-\varphi^{\rm SS} \left(-\bm{k},-i\varepsilon_{n} \right), 
\label{phiss-1band}
\end{eqnarray}
where 
$\varphi^{\rm SS} \left( \bm{k}, i\varepsilon_{n} \right)= \varphi^{\rm SS} \left(-\bm{k}, i\varepsilon_{n} \right)$ means the ESE state, and 
$\varphi^{\rm SS} \left( \bm{k}, i\varepsilon_{n} \right)=-\varphi^{\rm SS} \left(-\bm{k},-i\varepsilon_{n} \right)$ means the OSO state. 
For spin-triplet pairing, the gap function is 
\begin{eqnarray}
   \varphi^{\rm ST} \left( \bm{k}, i\varepsilon_{n} \right) 
 =-\varphi^{\rm ST} \left(-\bm{k}, i\varepsilon_{n} \right)
 = \varphi^{\rm ST} \left(-\bm{k},-i\varepsilon_{n} \right), 
\label{phist-1band}
\end{eqnarray}
where 
$\varphi^{\rm ST} \left( \bm{k}, i\varepsilon_{n} \right)=-\varphi^{\rm ST} \left(-\bm{k}, i\varepsilon_{n} \right)$ means the ETO state, and 
$\varphi^{\rm ST} \left( \bm{k}, i\varepsilon_{n} \right)= \varphi^{\rm ST} \left(-\bm{k},-i\varepsilon_{n} \right)$ means the OTE state.

\begin{figure}[!htb]
\begin{center}
\includegraphics[width=8.4cm]{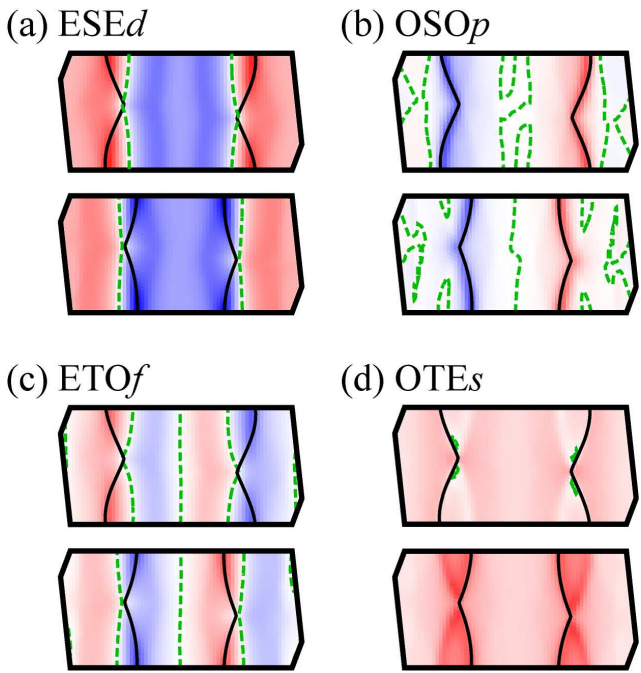}
\end{center}
\caption{
Gap functions of the outer (upper panels) and inner (lower panels) bands for (a) ESE and (b) OSO states. 
The triplet-pairing gap functions of the outer (upper panels) and inner (lower panels) bands for the (c) ETO and (d) OTE states. 
Note that the black solid curves represent the FS, the green dashed curves are nodes of the gap function, and the red (blue) contours represent the positive (negative) SC gap function. 
}
\label{fig4} 
\end{figure}

In the spin-singlet channel, the even-frequency gap functions are shown in the upper (lower) panel of Fig. \ref{fig4}(a) for the outer (inner) FS, which shows that 
the sign of the SC gap does not change on intrabands but changes on interbands. 
Since the SC gap is almost fully opened and has a positive (negative) value at the outer (inner) FS, the gap function has even-parity symmetry in wavenumber space. 
We expect that the nodal positions are slightly shifted from the Q points, although this is a subtle problem. 
The gap functions for odd-frequency pairing are shown in Fig. \ref{fig4}(b); these change the sign of the SC gap between the left and right FSs because they must satisfy the odd-parity gap function.

Figures \ref{fig4}(c) and \ref{fig4}(d) show the even- and the odd-frequency gap functions for the spin-triplet channel. 
Because the gap function in the ETO state must have odd parity, the sign of the gap on the right FS is opposite to that of the gap on the left FS. 
The OTE gap functions indicate the full-gap nature in wavenumber space, although the sign changes in frequency space.

To understand the origin of the gap symmetries, we adopt a single-band discreption. 
The spin and charge susceptibilities in Eqs. (\ref{chisp-tpsc}) and (\ref{chich-tpsc}) are simplified as 
\begin{eqnarray}
 \chi^{\rm sp}\left( q \right)
  = \frac{\chi^{0}\left( q \right)}{1-\chi^{0}\left( q \right) U^{\rm sp}}, \,\,\,\,
 &&
 \chi^{\rm ch}\left( q \right)
  = \frac{\chi^{0}\left( q \right)}{1+\chi^{0}\left( q \right) U^{\rm ch}},
  \label{chisc-1band}
\end{eqnarray}
respectively.  
The bare susceptibility $\chi^{0}\left( q \right)$ takes large positive values near the nesting vector $\bm{Q}$ at the lowest frequency and increases with decreasing temperature. 
Because we only introduce the on-site interaction, the spin susceptibility increases near the nesting vector $\bm{Q}$, and the charge susceptibility is suppressed by increasing on-site interaction. 
Thus, the relation 
$\chi^{\rm sp}\left( \bm{Q} \right) \gg \chi^{\rm ch}\left( \bm{Q} \right)$ 
is satisfied in the positive on-site interaction regime.

The relation gives the pairing interactions in Eqs. (\ref{pairing_int_ss}) and (\ref{pairing_int_st}) as 
\begin{eqnarray}
 V^{\rm SS}\left( Q \right) 
  &\simeq&  \frac{3}{2} U^{\rm sp} \chi^{\rm sp}\left( Q \right) U \ge 0, 
  \label{pairing_int_ss-1band}
\\
 V^{\rm ST}\left( Q \right) 
  &\simeq& -\frac{1}{2} U^{\rm sp} \chi^{\rm sp}\left( Q \right) U \le 0, 
  \label{pairing_int_st-1band}
\end{eqnarray}
respectively. 
From Eq. (\ref{gap-eq}), the gap function and the pairing interaction near the nesting vector $\bm{Q}$ must satisfy the following sign relation: 
\begin{eqnarray}
 \varphi^{\mu} \left( k+Q \right) &\sim& 
  - V^{\mu}\left( Q \right) \varphi^{\mu} \left( k \right). 
\label{sign-gap-func}
\end{eqnarray}
In other words, the sign of the gap function changes (does not change) between scattering of the nesting vector because the sign of the pairing interaction for the spin singlet (triplet) state is positive (negative). 
From now on, we will call this relation nesting--gap-sign rule.

We adopt the four-band model, in which the band structure is folded along the $b$ direction in the crystal structure and four Fermi sheets exist in the irreducible Brillouin zone. 
In the two-band or simplified one-band Q1D models, the $d$-wave-like gap function, which changes sign four times along the FS, seems to dominate in the ESE channel, thus assuming an unconventional glue such as spin fluctuations 
\cite{Kuroki2006-JPSJ-Rev}. 
When a magnetic field is applied or charge fluctuations coexist with spin fluctuations, an $f$-wave-like gap function, which changes sign six times along the FS, can arise in the ETO channel 
\cite{Kuroki2006-JPSJ-Rev}. 
Although the FS of the model for (TMTSF)$_2$ClO$_4$ differs from that of the one- or two-band model because of band-structure folding, we adopt a similar notation based on the symmetry of the SC gap function in wavenumber space.

\begin{figure}[!htb]
\begin{center}
\includegraphics[width=8.4cm]{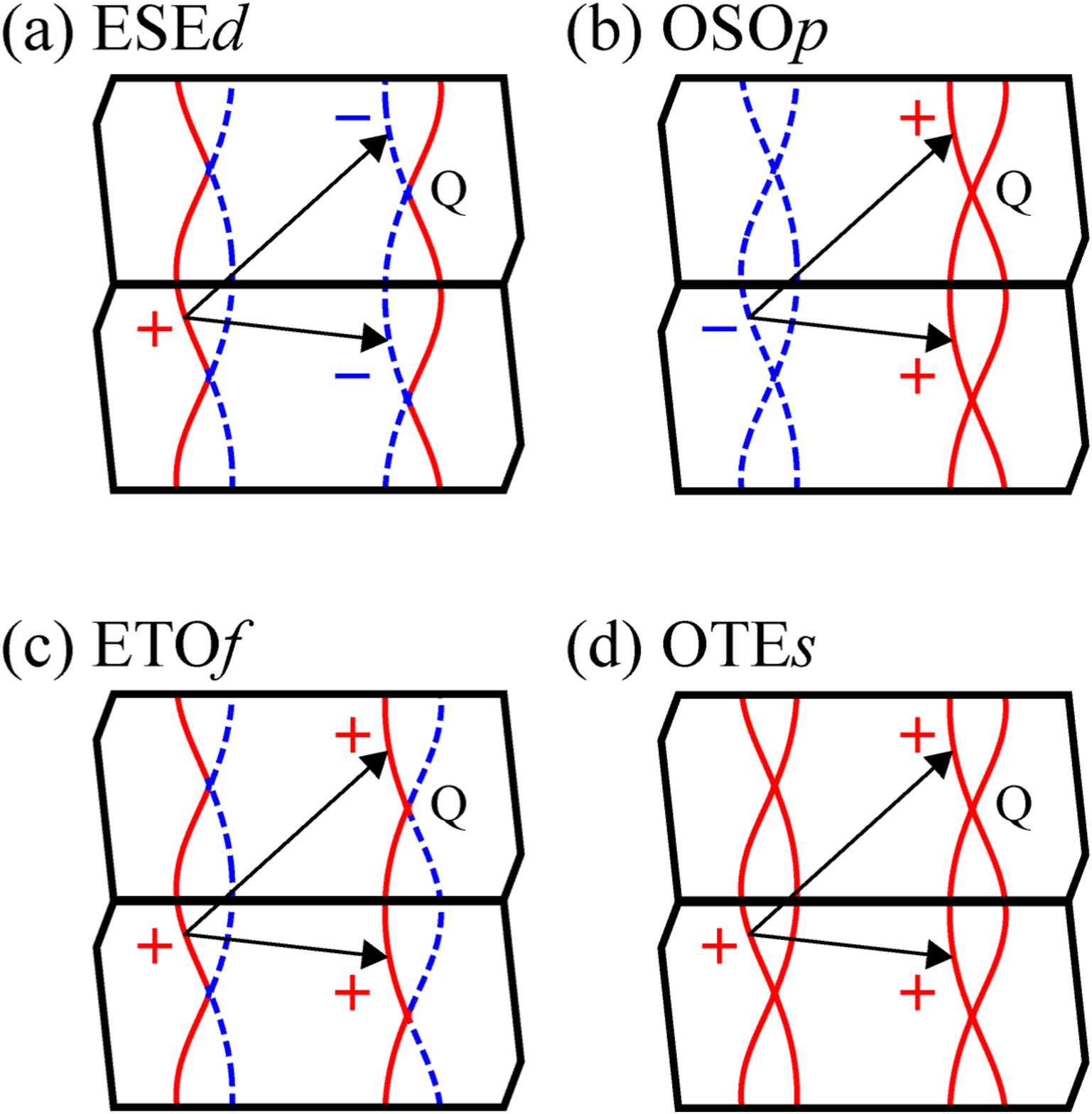}
\end{center}
\caption{
Sign of the gap function on FS based on results of Fig. \ref{fig4} for (a) ESE$d$, (b) OSO$p$, (c) ETO$f$, and (d) OTE$s$, where the red solid (blue dashed) curves represent the FS with positive (negative) sign of the SC gap, and the arrows represent the nesting vector $\bm{Q}_{2k_{\rm F}}$. 
"Q" represents the Q points.}
\label{fig5}
\end{figure}

Because we should focus on the sign of the gap function on the FS, the results in Fig. \ref{fig4} are represented in another form in Fig. \ref{fig5}. 
To highlight the effect of folding the band structure, we color the FS according to the sign of the SC gap in the irreducible Brillouin zone along the $b^{*}$ direction. 
In the ESE channel, as shown in Fig. \ref{fig5}(a), the $d$-wave gap function can be stabilized, although the gap may be regarded as an extended $s$-wave or $s_{\pm}$-wave, because the gap should satisfy (i) even parity and (ii) nesting--gap-sign rule. 
Figure \ref{fig5}(a) indicates that the sign of the gap changes between the interFSs, which is consistent with the nesting properties seen in Figs. \ref{fig3}(c) and \ref{fig3}(d). 
In the MLWO tight-binding model, the ESE$d$-SC gap is similar to the result in a simplified Q1D model with four FSs
\cite{Shimahara2000}. 
We discuss effects of $E_{\rm AO}$ on both the SC gap and pairing competition in Appendix \ref{appendix}. 
In the OSO channel shown in Fig. \ref{fig5}(b), the $p$-wave gap function with one irreducible nodal line between the left and right FSs can arise because of odd parity and nesting--gap-sign rule. 
Therefore, from now on, the ESE$d$ (OSO$p$) represents the $d$-wave ($p$-wave) gap in the ESE (OSO) pairing state.

In the ETO pairing shown in Fig. \ref{fig5}(c), the $f$-wave gap function, which has three nodes (adding one node between the left and right FS to the $d$-wave gap), can be stabilized by satisfying odd parity and nesting--gap-sign rule. 
In the OTE channel of Fig. \ref{fig5}(d), the spin-triplet $s$-wave fully-gapped function on the both FSs can appear. 
Thus, from now on, ETO$f$ (OTE$s$) represents the $f$-wave ($s$-wave) gap in the ETO (OTE) pairing state.

\subsection{Temperature dependence of eigenvalues}\label{ssec-res-temp-dep}

\begin{figure}[!htb]
\begin{center}
\includegraphics[width=8.4cm]{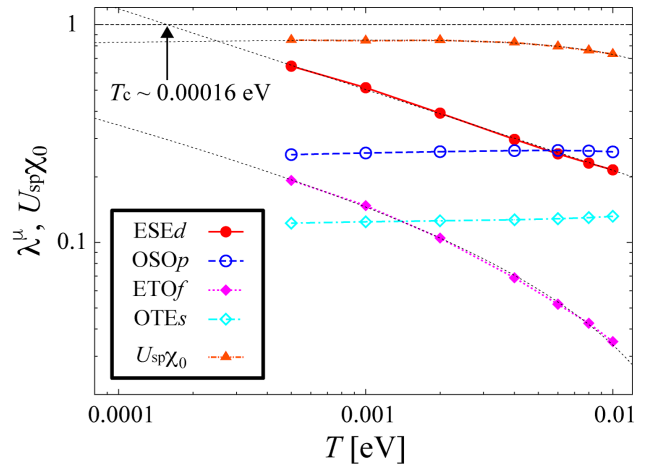}
\end{center}
\caption{
Temperature dependence of eigenvalue of $U_{\rm sp} \chi_{0}$ and $\lambda^{\mu}$, where the orange dash-dotted curve with solid triangles represents $U_{\rm sp} \chi_{0}$, the red solid (blue dashed) curve with solid (open) circles represents $\lambda^{{\rm ESE}d}$ ($\lambda^{{\rm OSO}p}$), the purple dotted (light-blue dash-dotted) curve with solid (open) diamonds represents $\lambda^{{\rm ETO}f}$ ($\lambda^{{\rm OTE}s}$), and the thin black dotted curves are extrapolations. 
}
\label{fig6}
\end{figure}

Based on the structure of the spin susceptibility and the SC gap functions, we show the temperature dependence of the eigenvalue of $U_{\rm sp} \chi_{0}$ for the magnetic order and that of the Eliashberg equation $\lambda^{\mu}$ for the SC order, where $\mu$ represents the pairing state. 
The temperature at which the eigenvalue reaches unity gives the critical temperature for its ordered state. 
The eigenvalue of $U_{\rm sp} \chi_{0}$ increases when the temperature is reduced, then saturates around 0.004 eV, as shown in Fig. \ref{fig6}. 
We do not consider this suppression caused by the Mermin--Wagner theorem, because according to this theorem, the eigenvalue of $U_{\rm sp} \chi_{0}$ asymptotically approaches unity if the SDW state is suppressed \cite{Aizawa2015}. 
The eigenvalue of $U_{\rm sp} \chi_{0}$ saturates below unity at lower temperature, as shown in Fig. \ref{fig6}, which suggests that the SDW state cannot arise at low temperature for the model used herein for (TMTSF)$_2$ClO$_4$. 
Our result for the absence of the 2$k_{\rm F}$-SDW in the model of slow-cooled (TMTSF)$_2$ClO$_4$ is consistent with the experimental observation, although the antiferromagnetic correlation has been found to develop in the low temperature regime in the NMR experiment
\cite{Bourbonnais1984,Shinagawa2007}.

As for the pairing competition, lowering the temperature leads to the development of the SC in the ESE$d$ and ETO$f$ states (see Fig. \ref{fig6}). 
In the competition between the ESE$d$ and ETO$f$, the eigenvalue $\lambda^{{\rm ESE}d}$ is greater than that of the ETO$f$. 
Upon extrapolation of the temperature dependence of $\lambda^{{\rm ESE}d}$, the SC occurs at about $T_{c}=1.6 \times 10^{-4}$ eV $\sim$ 1.9 K which is similar to the result obtained from experiments. 
As indicated by Eqs. (\ref{chisc-1band})--(\ref{pairing_int_st-1band}), the spin fluctuation contributes mainly to the pairing interaction. 
Then, based on Eqs. (\ref{pairing_int_ss-1band}) and (\ref{pairing_int_st-1band}), the absolute value of the pairing interaction for spin-singlet pairing is three times larger than that for spin-triplet pairing. 
Thus, we conclude that the ESE$d$ state is dominant and gives the critical temperature.

Conversely, the odd-frequency pairings, such as OSO$p$ and OTE$s$, saturate  with decreasing temperature, as seen in Fig. \ref{fig6}. 
The reason for this involves the stabilization of odd-frequency pairing near the magnetic transition \cite{Fuseya2003,Shigeta2012}. 
Although we do not exclude the possibility of odd-frequency pairing in this material, odd-frequency pairings are difficult to obtain at least in the Hubbard model with the parameter set used herein, since the tendency towards SDW ordering saturates at low temperatures.

\subsection{Fermi velocity and nodal position}\label{ssec-res-vf}

We focus on the position of the nodes on the FS for the ESE$d$ state. 
The recent angle-resolved heat capacity measurement suggests that the nodes of the gap appear around a wavenumber vector for which the angle between the $a$ axis and the Fermi velocity, $\phi_{{\bm v}_{\rm F}}$, is $\pm 10^{\circ}$\cite{Yonezawa2012}.

\begin{figure}[!htb]
\begin{center}
\includegraphics[width=7.4cm]{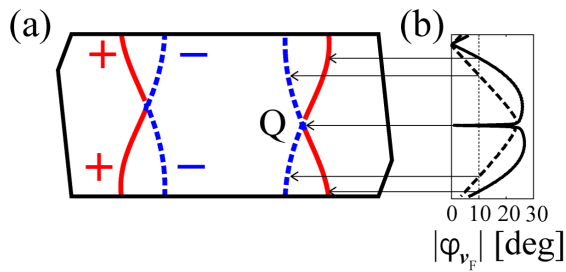}
\end{center}
\caption{
(a) FS colored by the sign of the ESE$d$-SC gap, where the red (blue) curves are for the positive (negative) sign of the gap. 
(b) the dependence of $\left| \phi_{{\bm v}_{\rm F}} \right|$ on $k_{y}$, the solid (dashed) curves are obtained from the outer (inner) FS, and the arrows show the ${\bm k}$ points that satisfy $\left| \phi_{{\bm v}_{\rm F}} \right| = 10^{\circ}$. 
}
\label{fig7} 
\end{figure}

Figure \ref{fig7}(a) shows the FS colored by the sign of the SC gap from Fig. \ref{fig5}(a). 
Figure \ref{fig7}(b) shows the $k_{y}$ dependence of $\left| \phi_{{\bm v}_{\rm F}} \right|$. 
We investigate that five ${\bm k}$ points satisfy the condition $\left| \phi_{{\bm v}_{\rm F}} \right| \simeq 10^{\circ}$. 
Among these ${\bm k}$ points, the node of the gap appears at the Q points. 
The angle $\left| \phi_{{\bm v}_{\rm F}} \right|$, however, can range from 0$^{\circ}$ to 20$^{\circ}$ around this ${\bm k}$ point because the FS varies discontinuously. 
This range includes the angle suggested in ref. \cite{Yonezawa2012}, although the error bar is large. 
The gap minima appearing at the Q points is also consistent with the experimental observation in ref. \cite{Pratt2013}.

\section{Conclusion}\label{ssec-conclusion}

We have performed first-principles band calculations for the Q1D organic superconductor (TMTSF)$_{2}$ClO$_{4}$. 
The outer FS almost touches the inner FS because of the small AO potential. 
Four band structures near the Fermi level are isolated from the others. 
Regarding the four band structures as the target bands, we derive an effective tight-binding model by applying the MLWO.

By applying the TPSC method to the multisite Hubbard model of (TMTSF)$_{2}$ClO$_{4}$, we calculate the spin susceptibility and gap function under the assumption that the pairing mechanism is mediated by spin fluctuations. 
The spin susceptibility is maximized at the nesting vector whose $k_{x}$ component has wavenumber $2k_{\rm F}$. 
The nesting vector $\bm{Q}_{2k_{\rm F}}$ connects the 
interFSs.

We obtain the eigenvalue and gap function of the linearized Eliashberg equation for the ESE$d$, OSO$p$, ETO$f$, and OTE$s$ channels. 
We find that the ESE$d$ pairing is dominant and its gap sign changes between the interFSs. 
With the on-site interaction $U$ adopted in this study and by applying the TPSC scheme, the critical temperature is estimated to be about 1.6$\times$10$^{-4}$ eV $\sim$ 1.9 K in the ESE$d$ pairing which is consistent with the experiment.

To compare with the recent experiment for the nodal position, we investigate how the angle $\left| \phi_{{\bm v}_{\rm F}} \right|$ of the Fermi velocity depends on $k_{y}$. 
Although the allowable range of $\left| \phi_{{\bm v}_{\rm F}} \right|$ is large, the nodes obtained from this study satisfy the condition $\left| \phi_{{\bm v}_{\rm F}} \right| \simeq 10^{\circ}$, which is consistent with the recent measurement\cite{Yonezawa2012}. 
The gap minima appearing at the Q points is also consistent with the experiment\cite{Pratt2013}.

In this study, we ignore long-range electron--electron interactions and electron--phonon interaction. 
Previous studies on this group of materials, however, pointed out that these interactions are important for competition between the pairing states
\cite{Kohmoto2001,Kuroki2001,Onari2004,Tanaka2004,Fuseya2005,Kuroki2005,Nickel2005,Aizawa2008,Aizawa2009,Aizawa2009b,Shigeta2011,Mizuno2011,Shigeta2013}. 
Based on these previous studies, these interactions can favor both the spin-triplet and odd-frequency pairing in which the nodeless gap can appear. 
Several experiments have suggested that the SC gap in (TMTSF)$_{2}$ClO$_{4}$ is nodeless 
\cite{Belin1997,Pratt2013}. 
In a future study, we will analyze the pairing competition in a model that accurately reproduces the DFT band structure and, therefore, should provide more meaningful results. 
The SC gap obtained from our work is neither a simple full gap nor a nodal gap on the FS. 
It is hence an interesting problem to see whether the experiments for the thermodynamic quantities \cite{Belin1997,Yonezawa2012,Pratt2013,Jerome2016} can be understood by applying theoretical analyses based on the obtained SC gap.

\acknowledgments

We acknowledge S. Yonezawa for valuable discussions. 
This work is supported by Grants-in-Aid from the Yokohama Academic Foundation, the Science Research Promotion Fund from the Promotion and Mutual Aid Corporation for Private Schools of Japan, and the Sasakawa Scientific Research Grant from the Japan Science Society.

\appendix

\section{Effects of anion-ordering potential}
\label{appendix}

\subsection{Dependence of eigenvalues on $E_{\rm AO}$}\label{ssec-ao-def}

Although the first-principles calculation gives a small AO potential, here we consider the possibility that the AO potential is enhanced in the actual material, and regard the potential as a variable. 
Next, by using a microscopic approach, we verify the pairing-gap function and the effects of the separation between the interFSs. 

\begin{figure}[!htb]
\begin{center}
\includegraphics[width=8.6cm]{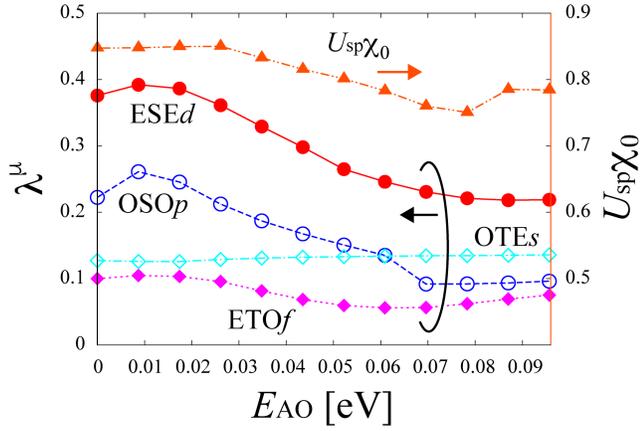}
\end{center}
\caption{
Eigenvalue of $U_{\rm sp} \chi_{0}$ and $\lambda^{\mu}$ as a function of the AO potential $E_{\rm AO}$, where the orange dash-dotted curve with solid triangles represents $U_{\rm sp} \chi_{0}$ (right axis), the red solid (blue dashed) curve with solid (open) circles represents $\lambda^{{\rm ESE}d}$ ($\lambda^{{\rm OSO}p}$), and the purple dotted (light-blue dash-dotted) curve with solid (open) diamonds represents $\lambda^{{\rm ETO}f}$ ($\lambda^{{\rm OTE}s}$). 
The eigenvalues of the SC state are plotted with respect to the left axis. 
}
\label{fig8} 
\end{figure}

Figure \ref{fig8} shows the eigenvalues of both $U_{\rm sp} \chi_{0}$ and the linearized Eliashberg equation as a function of the AO potential $E_{\rm AO}$ at $T=0.002$ eV. 
The eigenvalue of $U_{\rm sp} \chi_{0}$ decreases with increasing the AO potential, as shown by the right axis of Fig. \ref{fig8}, which suggests weakened spin susceptibility at the nesting vector $\bm{Q}_{2k_{\rm F}}$. 
When $E_{\rm AO}$ is greater than 0.08eV, $U_{\rm sp}\chi_{0}$ slightly increases again because the AO induces another nesting of the FS, which is discussed later. 
Conversely, $U_{\rm sp} \chi_{0}$ saturates when $E_{\rm AO}$ is smaller than 0.03eV, namely, the SDW may not develop in the limit of $E_{\rm AO} \to 0$ meV. 
This limit has been considered as corresponding to the experimental situation of fast cooling. 
We expect that the AO induced by slow cooling affects the transfer energies in addition to $E_{\rm AO} = 0$ meV, for example, the transfer energies $t_{S1}$ and $t_{S2}$ in the TMTSF chain A differs from them in the chain B as listed in TABLE \ref{transfers}. 
Assuming a spin-fluctuations-mediated pairing mechanism, increasing $E_{\rm AO}$ strongly suppresses the eigenvalues in the ESE$d$ and OSO$p$ states, compared with the eigenvalues in the triplet channels, as shown by both ETO$f$ and OTE$s$ (see the left axis of Fig. \ref{fig8}).

\subsection{Gap function and pairing mechanism under large anion-ordering potential}\label{ssec-res-gap-ao}

\begin{figure}[!htb]
\begin{center}
\includegraphics[width=8.4cm]{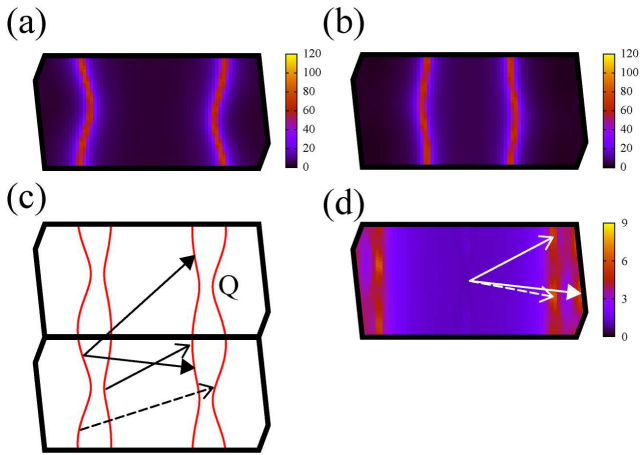}
\end{center}
\caption{
Green's function for (a) outer and (b) inner bands at $E_{\rm AO}=87$ meV. 
(c) FS and (d) the diagonalized spin susceptibility, where the arrows with the solid (open) point represent the nesting vector $\bm{Q}_{2k_{\rm F}}$ ($\bm{Q}_{\rm AO}$).} 
\label{fig9} 
\end{figure}

For an AO potential $E_{\rm AO}$=87 meV, which is 10 times greater than that obtained from the first-principles calculations, Fig. \ref{fig9}(a) [\ref{fig9}(b)] shows the absolute value of the Green's function for the outer (inner) band. 
Figure \ref{fig9}(c) shows the FS and the nesting vectors. 
Figure \ref{fig9}(d) shows the diagonalized spin susceptibility. 
The wavenumber vector that gives the maximum spin susceptibility occurs at $k_{x}$ slightly less than 2$k_{\mathrm F}$ \cite{Kishigi1997,Kishigi1998,Miyazaki1999a,Sengupta2001,Haddad2005,Haddad2007,Kishigi2007,Hasegawa2008}; we call this vector $\bm{Q}_{\rm AO}$. 
We now analyze the correspondence between $\bm{Q}_{\rm AO}$ and the nesting of the FS [see Fig. \ref{fig9}(c)]. 
The vector $\bm{Q}_{2k_{\rm F}}$ corresponds to the nesting vector of the interFSs which is similar to that shown in Fig. \ref{fig3}(c). 
The additional peaks around $\bm{Q}_{\rm AO}$ correspond to the nesting of the FS between the intrabands (intraFSs), as seen in Fig. \ref{fig9}(c). 
The nesting vector $\bm{Q}_{\rm AO}$ is originated from the separation of the FS induce by increasing $E_{\rm AO}$.

\begin{figure}[!htb]
\begin{center}
\includegraphics[width=8.4cm]{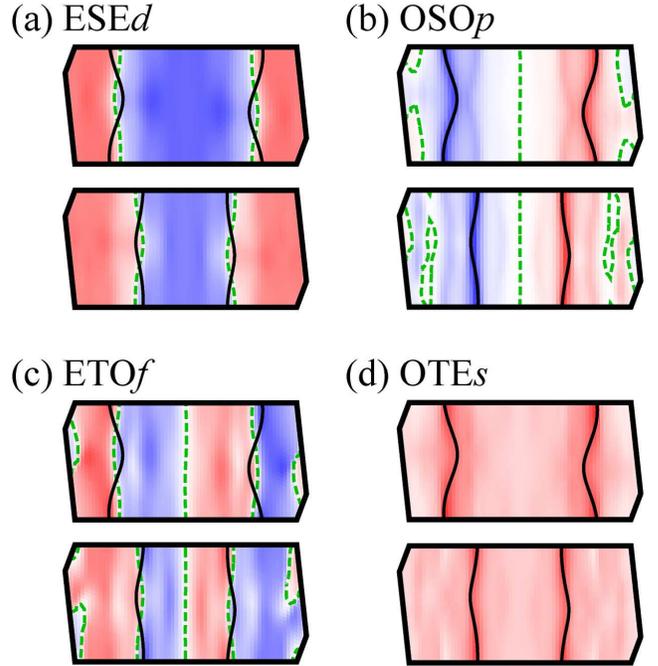}
\end{center}
\caption{
Gap functions of outer (upper panels) and inner (lower panels) bands for (a) the ESE$d$, (b) OSO$p$, (c) ETO$f$, and (d) OTE$s$ pairing states at $E_{\rm AO}$=87 meV. 
The black solid curves represent the FS, the green dashed curves are nodes of the gap function, and the red (blue) contour regime represents the positive (negative) SC gap function. 
}
\label{fig10} 
\end{figure}

Figure \ref{fig10}(a) shows the SC gap functions in the ESE channel. 
The nodal lines of the gap move onto the FSs because the nesting vector $\bm{Q}_{\rm AO}$ connects the intraFSs to satisfy the nesting--gap-sign rule for $\bm{Q}_{2k_{\rm F}}$ and $\bm{Q}_{\rm AO}$. 
The OSO$p$ gaps shown in Fig. \ref{fig10}(b) have no nodal line on the FS, thus producing the same gap as seen in Fig. \ref{fig4}(b). 
In the triplet channel, the gap functions require a constant sign between the nesting vectors by satisfying the nesting--gap-sign rule. 
Figure \ref{fig10}(c) shows the SC gap in the ETO$f$ state, which shows that the nodal lines move onto the FSs in a manner similar to that of the ESE$d$ state. 
The pairing gap of the OTE$s$ state also satisfies the sign relationship of the gap, as shown in Fig. \ref{fig10}(d).

\begin{figure}[!htb]
\begin{center}
\includegraphics[width=8.4cm]{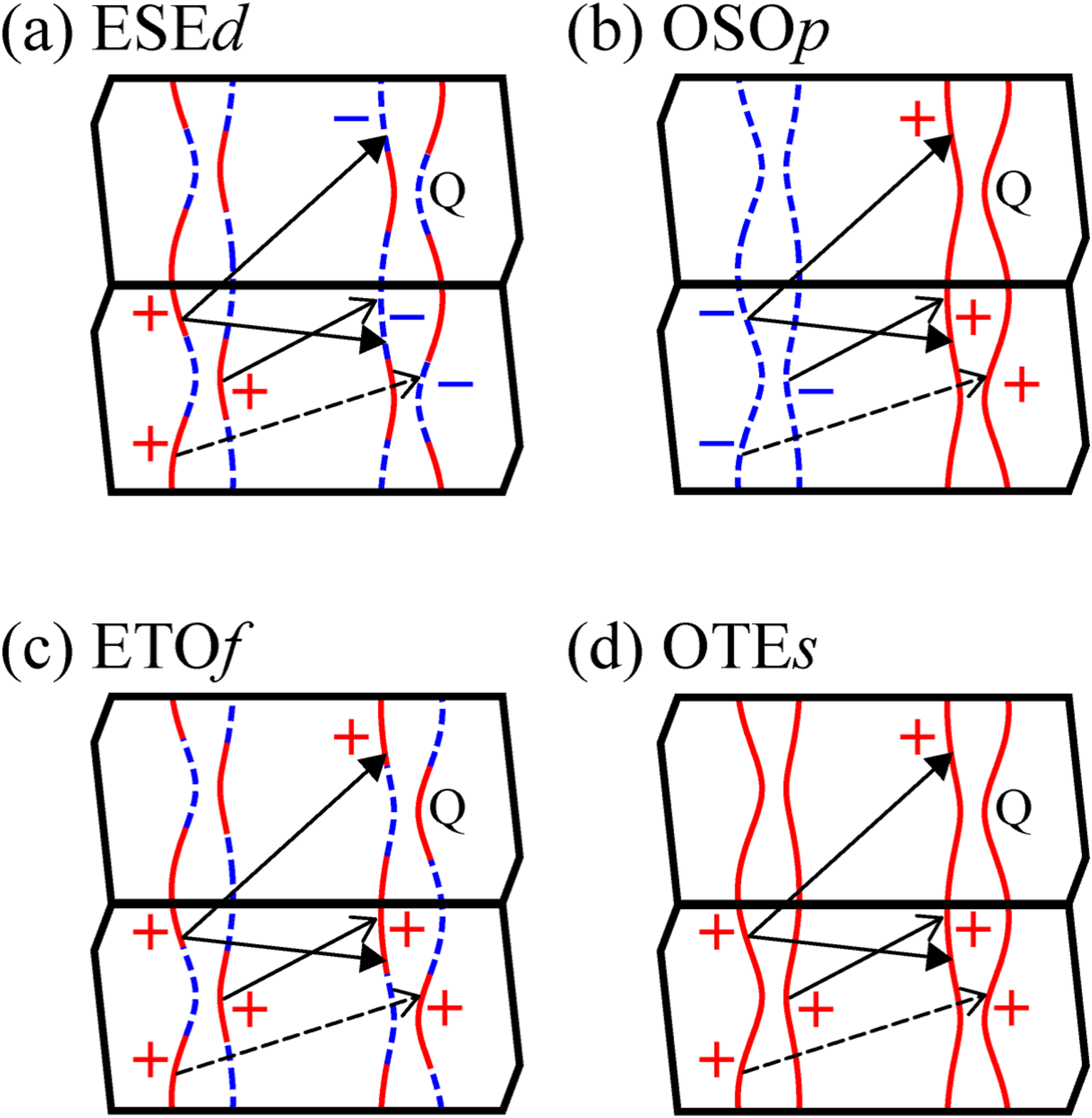}
\end{center}
\caption{
Sign of the gap function on the FS based on results of Fig. \ref{fig10} for (a) ESE$d$, (b) OSO$p$, (c) ETO$f$ and (d) OTE$s$, where the red solid (blue dashed) curves represent the positive (negative) sign of the pairing gap and the arrows with the solid (open) point represent the nesting vector $\bm{Q}_{2k_{\rm F}}$ ($\bm{Q}_{\rm AO}$). 
}
\label{fig11} 
\end{figure}

To consider effects of $E_{\rm AO}$ on the pairing mechanism, we focus on gap parity and nesting--gap-sign rule. 
For spin-singlet pairing, the gap function should satisfy even (odd) parity in even (odd) frequency. 
In the AO potential $E_{\rm AO}$ obtained from the first-principles calculation, $\bm{Q}_{2k_{\rm F}}$ that connects the interFSs is favorable. 
When parity of the gap function is satisfied, the $d$-wave-like gap in the extended zone can be favorable in the ESE channel, as shown in Fig. \ref{fig5}(a). 
When $E_{\rm AO}$ increases as a variable, the nesting vector $\bm{Q}_{\rm AO}$ that connects the intraFSs develops. 
Therefore, the gap functions require the nesting--gap-sign rule for $\bm{Q}_{2k_{\rm F}}$ and $\bm{Q}_{\rm AO}$. 
To satisfy parity and nesting--gap-rule, the ESE-SC gap needs additional nodes on the FSs, as shown in Fig. \ref{fig11}(a). 
Therefore, the nodeless $d$-wave SC gap 
\cite{Shimahara2000}, 
whose nodes are present between the separated FSs in the ESE channel, is unstable within this study. 
Although the gap function is similar to the $g$-wave-like gap function  suggested by Yonezawa \textit{et al.}\cite{Yonezawa2012}, we call this gap $d$-wave to be consistent with the preceding notation used herein. 
Conversely, the gap in the OSO channel needs no additional nodes [Fig. \ref{fig11}(b)] because this gap function is related to the handedness (right or left) of the FS.

In spin-triplet pairing, the gap function needs to be odd (even) parity in the even- (odd-) frequency pairing. 
The ETO$f$ gap in Fig. \ref{fig5}(c) can satisfy parity and nesting--gap-sign rule for $\bm{Q}_{2k_{\rm F}}$, which connects the interFSs. 
When the nesting vector $\bm{Q}_{\rm AO}$ develops by increasing $E_{\rm AO}$, the gap sign should unchange upon scattering by adding $\bm{Q}_{\rm AO}$ to $\bm{Q}_{2k_{\rm F}}$. 
To satisfy parity and nesting--gap-sign rule, the additional nodes come to the ETO-SC gap shown in Fig. \ref{fig11}(c). 
In the OTE channel, the gap function requires no additional nodes [see Fig. \ref{fig11}(d)] because this gap function does not depend on whether the FS is outer or inner.

We conclude that the SC is suppressed upon increasing $E_{\rm AO}$, from the viewpoint of (i) the strength of the pairing interaction and (ii) the addition of the nodes on the FS. 
When the pairing interaction mediated by the 2$k_{\rm F}$-spin fluctuations is weakened and even if the glue mediated by the $\bm{Q}_{\rm AO}$-spin fluctuations develops, the additional nodes appear on the FS and lead to suppression of SC. 

\begin{figure}[!htb]
\begin{center}
\includegraphics[width=7.4cm]{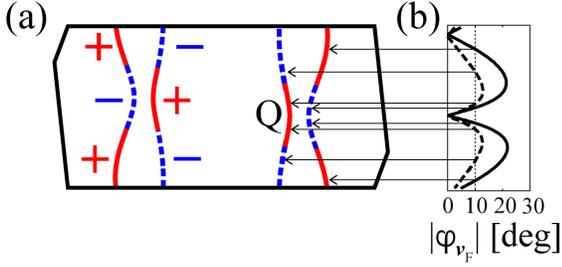}
\end{center}
\caption{
(a) FS colored by the sign of the ESE$d$ SC gap, where the red (blue) curves are for the positive (negative) sign of the gap. 
(b) The dependence of $\left| \phi_{{\bm v}_{\rm F}} \right|$ on $k_{y}$, the solid (dashed) curves are obtained from the outer (inner) FS, and the arrows show the ${\bm k}$ points that satisfy $\left| \phi_{{\bm v}_{\rm F}} \right| = 10^{\circ}$. 
}
\label{fig12} 
\end{figure}

Figure \ref{fig12}(a) shows the FS colored by the sign of the ESE$d$ gap from Fig. \ref{fig11}(a). 
Figure \ref{fig12}(b) shows the $k_{y}$ dependence of $\left| \phi_{{\bm v}_{\rm F}} \right|$ when $E_{\rm AO}$=87 meV. 
In Fig. \ref{fig12}(b), eight ${\bm k}$ points satisfy the condition $\left| \phi_{{\bm v}_{\rm F}} \right| \simeq 10^{\circ}$. 
Among these ${\bm k}$ points, although no nodes appear at the ${\bm k}$ points of $\left| \phi_{{\bm v}_{\rm F}} \right| \simeq 10^{\circ}$, the part of the FS close to the Q points has the gap nodes close to the ${\bm k}$ points of $\left| \phi_{{\bm v}_{\rm F}} \right| \simeq 10^{\circ}$.

\bibliography{text}

\end{document}